\documentclass[10pt,aps,pra,twoside]{revtex4}
\usepackage{epsf}
\begin{document}
\title{Quantum algorithms with fixed points:\\
       The case of database search}
\author{Lov K. Grover}
\email{lkgrover@bell-labs.com}
\affiliation{Bell Laboratories, Lucent Technologies,
             Murray Hill, NJ 07974, USA}
\author{Apoorva Patel}
\email{adpatel@cts.iisc.ernet.in}
\affiliation{Centre for High Energy Physics,
             Indian Institute of Science, Bangalore-560012, India}
\author{Tathagat Tulsi}
\email{tathagat@physics.iisc.ernet.in}
\affiliation{Department of Physics,
             Indian Institute of Science, Bangalore-560012, India}

\begin{abstract}
The standard quantum search algorithm lacks a feature, enjoyed by many
classical algorithms, of having a fixed-point, i.e. a monotonic convergence
towards the solution. Here we present two variations of the quantum search
algorithm, which get around this limitation. The first replaces selective
inversions in the algorithm by selective phase shifts of $\frac{\pi}{3}$.
The second controls the selective inversion operations using two ancilla
qubits, and irreversible measurement operations on the ancilla qubits
drive the starting state towards the target state. Using $q$ oracle queries,
these variations reduce the probability of finding a non-target state from
$\epsilon$ to $\epsilon^{2q+1}$, which is asymptotically optimal. Similar
ideas can lead to robust quantum algorithms, and provide conceptually new
schemes for error correction.

\end{abstract}
\maketitle

\section{Introduction}

\begin{quote}
{\it The quantum search algorithm is like baking a souffle \ldots you have
to stop at just the right time or else it gets burnt.~\cite{brass_science}}
\end{quote}

The quantum search algorithm~\cite{grover96} consists of an iterative sequence
of selective inversion and diffusion type operations. Each iteration results
in a fixed rotation (which is a function of the initial error probability)
of the quantum state in a two-dimensional Hilbert space formed by the source
and the target states. The iterative procedure keeps the rotation going
forever at a uniform rate. If we choose the right number of iteration steps,
we stop very close to the target state, else we keep on going round and round
in the two-dimensional Hilbert space. To perform optimally, therefore, we need
to know the precise number of iteration steps, which depends upon the initial
error probability or equivalently the fraction of target states in the
database. When we have this information, the quantum algorithm leads to a
square-root speed up over the corresponding classical algorithm for many
applications, including unsorted database search. When this information is
missing, we can estimate the required number of iteration steps using various
``amplitude estimation" algorithms~\cite{bhmt,grover98}, but that necessitates
an overhead of additional queries. When the total number of queries is large,
the additional queries do not cost much, but when the total number of queries
is small, the overhead can be unacceptably large.

In this article, we address the problem of finding an optimal quantum search
algorithm in situations where, (i) we do not know the initial error probability
(perhaps only its distribution or a bound is known), and (ii) the expected
number of queries is small (so that every additional query is a substantial
overhead). Such situations occur in pattern recognition and image analysis
problems (where each query is a significant cost), and in problems of error
correction and associative memory recall (where the initial error probability
is small but unknown). We look for variations of the quantum search algorithm
that ensure amplitude enhancement, and also outperform classical search
algorithms.

The strategy is familiar from classical computation, i.e. construct a
quantum algorithm that ``converges" towards the target state. That is,
however, impossible to do by iterating a non-trivial unitary transformation;
eigenvalues of unitary transformations are of magnitude 1, so the best that
can be achieved by iterating them is a ``limit cycle" and not a ``fixed point".
As described above, this is indeed what happens in case of the quantum search
algorithm. To obtain an algorithm that converges towards a fixed point,
some new ingredient is needed, and several possibilities come to mind:
\newline (a) Some property of the current computational state offers an
estimate of the distance to the target state. This estimate can be used as
a parameter to control the extent of the next iterative transformation,
e.g. the Newton-Raphson method to find the zeroes of a function. If such an
estimate is not available, we have to look for a different method.
\newline(b) Suitably designed but distinct operations are performed at
successive iterations. Such a method can converge towards a fixed point
even using unitary transformations. We illustrate this method using the
Phase-$\pi/3$ search algorithm~\cite{pi3}, described in detail in Section II.
\newline(c) Irreversible damping is introduced in the algorithm without
explicit use of any property of the target state. With the right type of
irreversibility (i.e. when all eigenvalues of the fixed iterative
transformation are less than 1 in magnitude), the algorithm converges,
e.g. the Gauss-Seidel method for solving a set of linear algebraic equations.
Within the framework of quantum computation, such an irreversibility can be
introduced by projective measurement operations, and we provide such an
algorithm in Section III~\cite{tath}.

In a broader context, fixed point algorithms possess two attractive features
by construction, which are not commonplace in generic quantum algorithms:
\newline (1) The initial state is guaranteed to evolve towards the target
state, even when the algorithm is not run to its full completion.
\newline (2) Any errors due to imperfect transformations in earlier
iterations are wiped out by the subsequent iterations, as long as the
state remains in the problem defining space.
\newline These are powerful motivations to explore quantum fixed point
algorithms, even if they have other limitations.

Let us consider an unsorted database in which a fraction $f$ of items are
marked, but we don't have precise knowledge of $f$. We run a particular
algorithm which has to return a single item from the database. If the returned
item is a marked one, the algorithm has succeeded, otherwise it is in error.
Without applying any algorithm, if we pick an item at random, then the
probability of error is $\epsilon=1-f$. The goal of the algorithm is to
minimize the error probability, using the smallest number of oracle queries.
If $f$ is sufficiently small, then we can use the optimal quantum search
algorithm to obtain a marked item using $O(1/\sqrt{f})$ queries. A few more
queries to estimate $f$ or to fine-tune the algorithm is not a problem,
because overall we gain a quadratic speed-up compared to the classical case
requiring exhaustive search. But when $f$ is large, the number of oracle
queries is small, and the quantum search algorithm doesn't provide much
advantage---in particular the rotation can overshoot the target state.
In such a situation, a simple classical algorithm (select a random item
and use a query to check if it is a marked one) may outperform the quantum
search algorithm. The same considerations apply to the more general amplitude
amplification algorithms~\cite{bhmt,grover98}.
There the initial quantum state is a unitary operator $U$ applied to a given
source state $|{s}\rangle$, and the probability of getting a target state
after measuring this initial state, $|U_{ts}|^{2}$, is analogous to $f$.
The probability of error, which has to be minimized using the smallest number
of queries, is the probability of getting a non-target state after measurement,
$\epsilon=1-|U_{ts}|^2$.

In Section II, we show that by replacing the selective phase inversions in
quantum search by suitable phase shifts we can get an algorithm that always
produces amplitude amplification. As shown in Fig.1, the change in phase
shift eliminates overshooting and the state vector \textsl{always} moves
closer to the target state. By recursively applying the single iteration
derived for any unitary operator $U$, we develop an algorithm with multiple
applications of $U$ that converges monotonically to the target state.
The optimal value of the phase shift turns out to be $\pi/3$, and we refer
to this algorithm as the ``Phase-$\pi/3$ search"~\cite{pi3}.
Remarkably, the algorithm converges using reversible unitary transformations
and without ever estimating the distance of the current state from the target
state. Explicitly, the initial error probability $\epsilon$ changes to
$\epsilon^3$ in one iteration. Recursive application of the basic iteration
$n$ times requires $q_{i}=3q_{i-1}+1$ ($q_{0}=0$) oracle queries at the
$i^{\mathrm{th}}$ level, and makes the error probability $\epsilon^{3^n}$.
Thus $\epsilon=0$ is the fixed point of the algorithm, and the error
probability decreases as $\epsilon^{2q+1}$ as a function of the number of
queries $q$.

Section III describes a different implementation of the fixed point quantum
search, where irreversible measurement operations direct the current state
towards the target state~\cite{tath}. In it the inversion and diffusion
operations are controlled in a special way by two ancilla qubits and their
measurement. The same transformation is repeated at every iteration, but
since the transformation is made non-unitary by measurement, the quantum
state is able to monotonically converge towards the target state. A major
advantage of this implementation is that the $\epsilon^{2q+1}$ convergence
is achieved for all positive integer values of $q$, compared to the
restricted numbers $q_n=(3^n-1)/2$ for the Phase-$\pi/3$ search.

Note that the best classical algorithm can only decrease the error probability
as $\epsilon^{q+1}$ (not $\epsilon^{q}$, since the last iteration does not
need a query). Thus the fixed point quantum search improves the convergence
rate by a factor of 2. In Section IV, we illustrate this difference with an
explicit example.

\begin{figure}[ptb]
\begin{center}
\begin{picture}(500,100)(0,0)
\put(120,15){\vector(1,0){70}}
\put(120,15){\vector(0,1){70}}
\put(120,15){\vector(1,1){53}}
\put(120,15){\vector(-1,1){53}}
\put(120,90){\makebox(0,0)[r]{$\vert{t}\rangle$}}
\put(205,12){\makebox(0,0)[tr]{$\vert{t_\perp}\rangle$}}
\put(170,77){\makebox(0,0)[tl]{$U\vert{s}\rangle$}}
\put(45,77){\makebox(0,0)[tl]{$U I_s U^\dagger I_t U\vert{s}\rangle$}}
\put(280,15){\vector(1,0){70}}
\put(280,15){\vector(0,1){70}}
\put(280,15){\vector(1,1){53}}
\put(280,15){\vector(1,3){22}}
\put(280,90){\makebox(0,0)[r]{$\vert{t}\rangle$}}
\put(365,12){\makebox(0,0)[tr]{$\vert{t_\perp}\rangle$}}
\put(330,77){\makebox(0,0)[tl]{$U\vert{s}\rangle$}}
\put(290,92){\makebox(0,0)[tl]{$U R_s U^\dagger R_t U\vert{s}\rangle$}}
\end{picture}
\caption{In the standard quantum search algorithm (left), the state vector
may overshoot the target state. In the algorithm of Section II (right), the
state vector always moves towards the target state.}
\end{center}
\end{figure}
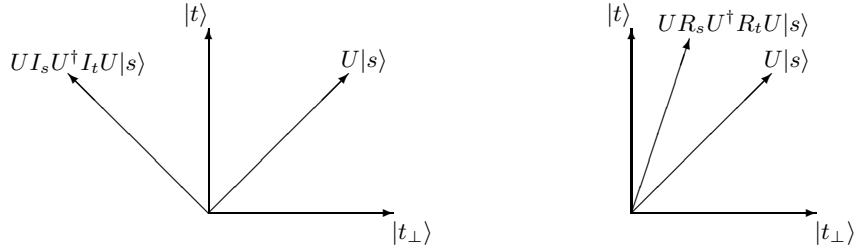

\section{The Phase-$\pi/3$ Search}

Consider applying the transformation
\begin{equation}
O(\theta,\phi) =  UR_{s}^{\theta}U^{\dagger}R_{t}^{\phi}U ~,
\end{equation}
to the state $|{s}\rangle$, where $R_{s}^{\theta}$ and $R_{t}^{\phi}$
are selective phase shift operators for the source and the target state
respectively. Note that if we were to choose the phase shifts as
$\theta=\phi=\pi$, we would get one iteration of the amplitude
amplification algorithm~\cite{bhmt,grover98}.

In the following, we study the particular case $\theta=\phi=\pi/3$.
We show that when the $U$ operation drives the state vector from $|s\rangle$
to $|t\rangle$ with a probability $(1-\epsilon)$, i.e. $\vert U_{ts} \vert^2
=(1-\epsilon)$, then $O(\pi/3,\pi/3)$ drives the state vector from the source
state to the same target state with a probability of $(1-\epsilon^3).$
The deviation from $|t\rangle$ hence falls from $\epsilon$ to $\epsilon^3$.
The striking aspect of this result is that it holds for \textsl{any} kind of
deviation from $|t\rangle$. Unlike the standard quantum search algorithm,
which would overshoot the target state when $\epsilon$ is small (Fig.1),
the Phase-$\pi/3$ search always moves towards the target state. This property
is useful in developing quantum algorithms that are robust to variations in
the problem parameters.

Connections to error correction are already evident in the previous paragraph.
Let us say that we want to drive a system from $|s\rangle$ to $|t\rangle$
state/subspace, using the transformation $U$, with driving probability
$\vert U_{ts}\vert^2 = (1-\epsilon)$. Then the composite transformation
$O(\pi/3,\pi/3)$ will reduce the error from $\epsilon$ to $\epsilon^3$.
This technique is applicable whenever the transformations $U$, $U^{\dagger}$,
$R_s$ and $R_t$ can be implemented. That will be the case when the errors
are either systematic or slowly varying, and the transformation $U$ can
be inverted with exactly the same error, e.g. when the errors are due to
environmental degradation of some component. The technique would not apply
to errors that arise as a result of sudden disturbances from the environment.
Quantum error correction is carried out at the single qubit level in many
implementations, where individual errors are corrected by identifying the
corresponding error syndrome. With the machinery of this paper, errors can
be corrected without ever needing to identify the error syndrome.

\subsection{Analysis}

Now let us analyze the effect of the transformation $O(\pi/3,\pi/3)$ when
it is applied to the state $|s\rangle$. We want to show that
\begin{equation}
\Big\vert\langle t| UR_{s}^{\pi/3}U^{\dagger}R_{t}^{\pi/3}U
|s \rangle\Big\vert^{2} = 1-\epsilon^{3} ~.
\end{equation}
Applying the operations, $U$, $R_s^{\pi/3}$, $U^\dagger$, $R_t^{\pi/3}$ and
$U$, one by one, to the state $|s\rangle$, we find that
\begin{equation}
O(\pi/3,\pi/3) |s\rangle = U|s\rangle \left( e^{i\pi/3} +\vert U_{ts} \vert^2
\left(e^{i\pi/3}-1\right)^{2} \right) + |t\rangle U_{ts}
\left(e^{i\pi/3}-1\right) ~.
\end{equation}
The deviation of this superposition from the target state $|t\rangle$
is given by the absolute square of its overlap with the non-target states.
This probability is
\begin{eqnarray}
\Big\vert \langle t_\perp| O(\pi/3,\pi/3) |s\rangle \Big\vert^2
&=& \langle s| O(\pi/3,\pi/3)^\dagger \Big( 1 - |t\rangle\langle t| \Big)
  O(\pi/3,\pi/3) |s\rangle \nonumber\\
&=& \left(  1-\left\vert U_{ts}\right\vert^2 \right)
  \left\vert \left( e^{i\pi/3} + \left\vert U_{ts}\right\vert^2
  \left( e^{i\pi/3}-1 \right)^2 \right) \right\vert^2 ~.
\end{eqnarray}
Substituting $\left\vert U_{ts}\right\vert^2 = (1-\epsilon)$,
the above quantity becomes
\begin{equation}
  \epsilon \left\vert \left( e^{i\pi/3}+\left(1-\epsilon\right)
  \left(e^{i\pi/3}-1\right)^2 \right) \right\vert^2
= \epsilon \left\vert \left( e^{2i\pi/3}-e^{i\pi/3}+1 \right)
  -\epsilon\left( e^{i\pi/3}-1 \right)^2 \right\vert^2
= \epsilon^{3} ~.
\end{equation}
We now use this result recursively to construct the quantum algorithm
for searching in presence of uncertainty.

\subsection{Recursion}

A few years after its invention, the quantum search algorithm
\cite{grover96,sch} was generalized to a much larger class of applications
known as the amplitude amplification algorithms~\cite{bhmt,grover98}.
In these algorithms, the amplitude to go from the state $|s\rangle$
to the state $|t\rangle$ by applying a unitary operation $U$, can be
\textsl{amplified} by repeating the sequence of operations,
$Q=I_{s}U^{\dag}I_{t}U$, where $I_{s}$ and $I_{t}$ denote selective
inversions of the $|s\rangle$ and $|t\rangle$ states respectively.
Note that the amplitude amplification transformation with four queries is
\begin{equation}
U\left( I_{s}U^{\dagger}I_{t}U\right) \left( I_{s}U^{\dagger}I_{t}U\right)
 \left( I_{s}U^{\dagger}I_{t}U\right) \left( I_{s}U^{\dagger}I_{t}U\right) ~.
\label{ampt-amp}%
\end{equation}
When the operation sequence $I_{s}U^{\dag}I_{t}U$ is repeated $\eta$ times,
the amplitude in the $U^{\dagger}|t\rangle$ state becomes approximately
2$\eta U_{ts}$ provided $\eta U_{ts}\ll 1$. The quantum search algorithm
is a particular case of amplitude amplification, with $U$ being the
Walsh-Hadamard transformation and $|s\rangle$ being the $|\overline{0}\rangle$
state with all qubits in the $|0\rangle$ state. The selective inversions
enable the amplitudes produced by successive iterations to add up in phase.
Consequently, the amount of amplification grows linearly with the number of
repetitions of $Q$, and the probability of obtaining $|t\rangle$ goes up
quadratically.

Just like the amplitude amplification transformation, it is possible to
recurse the transformation $UR_{s}U^{\dagger}R_{t}U$ to obtain larger
rotations of the state vector in a carefully defined two dimensional Hilbert
space. The basic idea is to define transformations $U_{i}$ by the recursion
\begin{equation}
U_{i+1}=U_{i}R_{s}U_{i}^{\dagger}R_{t}U_{i} ~,~~ U_{0} \equiv U ~.
\label{algorithm}
\end{equation}
Unlike amplitude amplification, there is no simple structure in the recursion,
and it is \textsl{not} easy to write down the actual operation sequence for
$U_{i}$ with large $i$, without working out the full recursion for all integers
less than $i$. Let us illustrate this for $U_2$:%
\begin{equation}
U_{0}=U ~,~~
U_{1}=U_{0}R_{s}U_{0}^{\dagger}R_{t}U_{0}=UR_{s}U^{\dagger}R_{t}U ~,
\end{equation}
\begin{eqnarray}
U_{2} = U_{1}R_{s}U_{1}^{\dagger}R_{t}U_{1}
&=& \left( UR_{s}U^{\dagger}R_{t}U\right) R_{s}
    \left( UR_{s}U^{\dagger}R_{t}U\right)^{\dagger} R_{t}
    \left( UR_{s}U^{\dagger}R_{t}U\right) \nonumber\\
&=& \left( UR_{s}U^{\dagger}R_{t}U\right) R_{s}
    \left( U^{\dagger}R_{t}^{\dagger}UR_{s}^{\dagger}U^{\dagger}\right) R_{t}
    \left( UR_{s}U^{\dagger}R_{t}U\right) \nonumber\\
&=& U\left( R_{s}U^{\dagger}R_{t}U\right)
     \left( R_{s}U^{\dagger}R_{t}^{\dagger}U\right)
     \left( R_{s}^{\dagger}U^{\dagger}R_{t}U\right)
     \left( R_{s}U^{\dagger}R_{t}U\right) ~.
\label{recursion3}
\end{eqnarray}
This is not as simple as the corresponding transformation for amplitude
amplification, Eq.(\ref{ampt-amp}).

Now it is straightforward to show that if $\vert U_{ts}\vert^2 = 1-\epsilon$,
then $\vert(U_i)_{ts}\vert^2 = 1-\epsilon^{3^i}$. The recursion obeyed by
the number of queries is, $q_{i+1} = 3q_i + 1$, $q_0=0$, with the solution
$q_i = (3^i-1)/2$. As a function of the number of queries, therefore,
$\vert (U_i)_{ts}\vert^2 = 1-\epsilon^{2q_i+1}$. The error probability
thus falls as $\epsilon^{2q+1}$ after $q$ queries, which is an improvement
over the classical algorithm where the error probability is $\epsilon^{q+1}$
after $q$ queries (more details in Section IV).

\subsection{Fixed Point of the Algorithm}

First, note that both the amplitude amplification algorithm,
Eq.(\ref{ampt-amp}), and the phase shift algorithm, Eq.(\ref{recursion3}),
perform selective operations for the state $|t\rangle$, and so from an
information theoretic point of view there is no conflict in having fixed
points. However, unitarity would be violated if there was accumulation of the
target state due to repetition of a fixed transformation. In the amplitude
amplification algorithm, exactly the same transformation is repeated and so
unitarity does not permit any fixed point. Although the phase shift algorithm
is very similar to the amplitude amplification algorithm, the transformation
applied at each step is slightly different from the others, due to the
presence of the four operations $R_{s},R_{t},R_{s}^{\dagger},R_{t}^{\dagger}$.
It therefore gets around the unitarity condition that prevents the amplitude
amplification algorithm from having a fixed point.

The $\epsilon^{2q+1}$ performance of the Phase-$\pi/3$ algorithm has been
shown to be asymptotically (i.e. in the limit $\epsilon\rightarrow0$)
optimal~\cite{jai}. It can be shown that application of the general
operator $O(\theta,\phi)$ to $|{s}\rangle$ changes its component along
$|t_\perp\rangle$ by the scale factor $\Big|\exp({i\over2}(\theta-\phi))
- 4\sin{\theta\over2}\sin{\phi\over2} |U_{ts}|^2\Big|$. In the asymptotic
limit, $|U_{ts}|\rightarrow1$, and the scale factor is minimized by the
choice $\theta=\phi={\pi\over3}$.

\section{Measurement Based Search Algorithm}

To obtain a fixed point quantum search, we have to find an algorithm that
successively decreases the probability of finding a non-target state.
Let us say that the initial state of a quantum register is $U\vert{s}\rangle
= \sin\alpha \vert{t}\rangle+ \cos\alpha\vert{t_{\perp}}\rangle$---an
unspecified superposition of the target $\vert t\rangle$ and the non-target
$\vert t_{\perp}\rangle$ states. We can construct a fixed point search
algorithm by repetitively measuring the state---whenever the oracle query
identifies the measured state as the target state, we stop the algorithm,
otherwise we use the diffusion operation to rotate the non-target state
towards the target state. To implement this idea, we attach to the quantum
register an ancilla bit in the initial state $\vert{0}\rangle$. The oracle
query flips the ancilla when the register is in the target state. Now if we
measure the ancilla, outcome $1$ tells us that we are done, and the register
will give us the target state. Outcome $0$ tells us that the register is
in a superposition $|t_\perp\rangle$ of the non-target states, and its
probability is the initial error probability $\epsilon=\cos^{2}\alpha$. To
decrease this probability, we apply the diffusion operator $UI_{s}U^{\dagger}$
to the register, conditioned on the measurement outcome being $0$. That
reflects $\vert{t_\perp}\rangle\vert{0}\rangle$ about $U\vert{s}\rangle
\vert{0}\rangle$ to give the state $\sin2\alpha\vert{t}\rangle\vert{0}\rangle
+ \cos2\alpha\vert{t_\perp}\rangle\vert{0}\rangle$. The error probability
has thus decreased by the factor $\cos^{2}2\alpha$. Iterating the sequence
of oracle query and diffusion operations $n$ times, the error probability is
reduced to $\cos^{2}\alpha \cos^{2n}2\alpha = \epsilon(2\epsilon-1)^{2n}$.
This convergence is better than the Phase-$\pi/3$ search when $\epsilon>1/3$,
but worse when $\epsilon<1/3$.

Our goal is to find an algorithm which gives optimal convergence for all
values of $\epsilon$, without knowing any bounds that $\epsilon$ may obey.
We know that the rotation provided by a single iteration of the quantum
search algorithm overshoots the target state when $\epsilon$ is small.
In particular, when $\epsilon$ is less than $1/2$, the overshooting is
so large that the new state has a smaller overlap with the target state
than the initial state. That gives an inkling for constructing a better
directed quantum search---somehow set a lower bound of $1/2$ for $\epsilon$.
The natural lower bound for $\epsilon$ is $0$, but we can easily make it
$1/2$ by diluting the database with extraneous states. Explicitly, we add
to the quantum register a second ancilla in the state $\vert{+}\rangle\equiv
(\vert{0}\rangle+\vert{1}\rangle)/\sqrt{2}$, and make the oracle query
conditional on this ancilla. This logic produces the following $q$-iteration
algorithm:

\begin{itemize}
\item Attach two ancilla qubits in the $|0\rangle$ state to the source
state $|s\rangle$, i.e. $|s\rangle\rightarrow|0\rangle|s\rangle|0\rangle$.
(In what follows, we refer to the former ancilla as ancilla-1 and the latter
ancilla as ancilla-2.)
\item Apply $H\otimes U\otimes I$ to this extended register to prepare the
initial state of the algorithm, $|+\rangle(U|s\rangle) |0\rangle$.
\item Iterate the following two steps, $q$ times:\newline
Step 1: If ancilla-1 is in the state $|1\rangle$, then perform an oracle
query that flips ancilla-2 when the register is in target state.\newline
Step 2: Measure ancilla-2. If the outcome is $1$, the register is certainly
in the target state, so exit the iteration loop. If the outcome is $0$, then
apply the joint diffusion operator $(H\otimes U)I_{0s}(H\otimes U)^{\dagger}$
to the joint state of ancilla-1 and the register.
\item After exiting or completing the iteration loop, stop the algorithm
and measure the register.
\end{itemize}

The quantum circuit for this algorithm is shown in Fig.2.

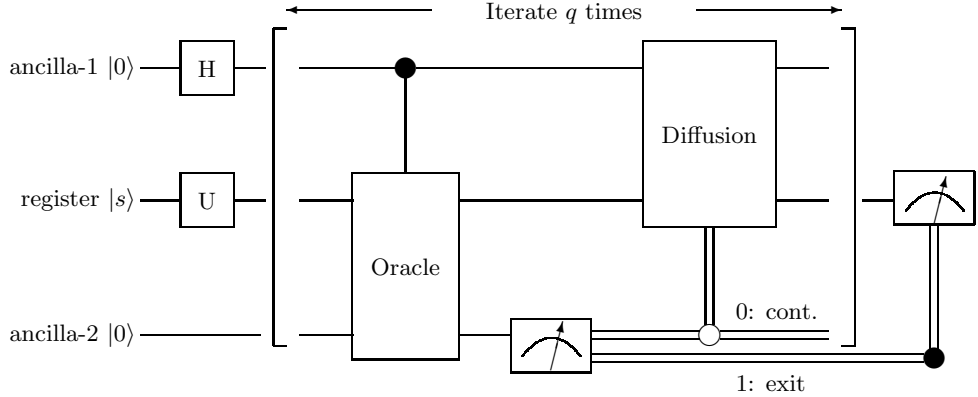
\begin{figure}[ptb]
\begin{picture}(300,160)(0,0)
\put(23,24){\makebox(0,0)[r]{ancilla-2 $\vert{0}\rangle$}}
\put(185,142){\makebox(0,0)[b]{Iterate $q$ times}}
\put(135,147){\vector(-1,0){55}}
\put(235,147){\vector(1,0){55}}
\put(25,24){\line(1,0){45}}
\put(25,75){\line(1,0){15}}
\put(25,125){\line(1,0){15}}
\put(40,65){\framebox(20,20){U}}
\put(40,115){\framebox(20,20){H}}
\put(60,75){\line(1,0){12}}
\put(60,125){\line(1,0){12}}
\put(85,24){\line(1,0){20}}
\put(85,75){\line(1,0){20}}
\put(105,15){\framebox(40,70){Oracle}}
\put(145,24){\line(1,0){20}}
\put(195,22.5){\line(1,0){41}}
\put(195,25.5){\line(1,0){41}}
\put(240,24){\circle{8}}
\put(244,22.5){\line(1,0){41}}
\put(244,25.5){\line(1,0){41}}
\put(250,30){$0$: cont.}
\put(238.5,28){\line(0,1){37}}
\put(241.5,28){\line(0,1){37}}
\put(215,65){\framebox(50,70){Diffusion}}
\put(145,75){\line(1,0){70}}
\put(265,75){\line(1,0){20}}
\put(295,20){\line(0,1){120}}
\put(75,20){\line(0,1){120}}
\put(290,20){\line(1,0){5}}
\put(290,140){\line(1,0){5}}
\put(75,20){\line(1,0){5}}
\put(75,140){\line(1,0){5}}
\put(23,125){\makebox(0,0)[r]{ ancilla-1 $\vert{0}\rangle$}}
\put(23,75){\makebox(0,0)[r]{ register $\vert{s}\rangle$}}
\put(85,125){\line(1,0){36}}
\put(125,125){\circle*{8}}
\put(129,125){\line(1,0){86}}
\put(125,121){\line(0,-1){36}}
\put(265,125){\line(1,0){20}}
\put(165,10){\framebox(30,20)}
\qbezier(169,16)(180,30)(191,16)
\put(180,11){\vector(1,4){4.5}}
\put(250,3){$1$: exit}
\put(195,17){\line(1,0){126}}
\put(195,14){\line(1,0){126}}
\put(325,15.5){\circle*{8}}
\put(323.5,19){\line(0,1){46}}
\put(326.5,19){\line(0,1){46}}
\put(298,75){\line(1,0){12}}
\put(310,66){\framebox(30,20)}
\qbezier(314,71)(325,85)(336,71)
\put(325,66){\vector(1,4){4.5}}
\end{picture}
\caption{Quantum circuit for our $q$-iteration algorithm.}
\end{figure}

\subsection{Analysis}

Let us analyze the algorithm step by step. The initial state is
\begin{equation}
\vert{\psi_{i}}\rangle= (H\otimes U\otimes I)\vert{0}\rangle\vert{s}
\rangle\vert{0}\rangle= \left( \frac{\vert{0}\rangle+ \vert{1}\rangle}
{\sqrt{2}}\right) \Big(\sin\alpha\vert{t}\rangle+\cos\alpha\vert{t_\perp}
\rangle\Big)\vert{0}\rangle~.
\end{equation}
The initial error probability is $\epsilon=\cos^2\alpha$, and the initial
success probability is $f=1-\epsilon=\sin^2\alpha$. We work in the joint
search space of ancilla-1 and the register, denoted by the subscript $j$,
where only the state $\vert{t_j}\rangle\equiv\vert{1}\rangle\vert{t}\rangle$
acts as the target state and all the states $\vert{0}\rangle\vert{t}\rangle,
\vert{0}\rangle\vert{t_\perp}\rangle, \vert{1}\rangle\vert{t_\perp}\rangle$
act as non-target states. Let $\vert{t^{\prime}_j}\rangle$ denote the
superposition of all the non-target states. In the joint search space,
the initial state is
\begin{equation}
\vert{\psi_i}\rangle= \left( \frac{\sin\alpha}{\sqrt{2}}\right) \vert{t_j}
\rangle\vert{0}\rangle+ \frac{1}{N}\vert{t^{\prime}_j}\rangle\vert{0}\rangle~,
\end{equation}
where the unit vector $|t^{\prime}_{j}\rangle$ is
\begin{equation}
\vert{t^{\prime}_j}\rangle= N\left( \frac{\sin\alpha}{\sqrt{2}}\vert
{0}\rangle\vert{t}\rangle+\frac{\cos\alpha}{\sqrt{2}}\vert{0}\rangle
\vert{t_\perp}\rangle+\frac{\cos\alpha}{\sqrt{2}}\vert{1}\rangle
\vert{t_\perp}\rangle\right)  ~,
\end{equation}
and the normalization constant $N$ is $\left( \cos^2\alpha + \frac{1}{2}
\sin^2\alpha \right)^{-1/2} = \sqrt{2/(1+\epsilon)}$. For later reference,
note that the error probability after measuring the joint state
$\vert{t^{\prime}_j}\rangle$, i.e. the probability of finding the register
in the non-target state $\vert{t_\perp}\rangle$, is
\begin{equation}
\vert\langle0 \vert\langle t_\perp\vert t^{\prime}_j\rangle\vert^2 +
\vert\langle1 \vert\langle t_\perp\vert t^{\prime}_j\rangle\vert^2 =
N^{2}\cos^2\alpha= N^{2}\epsilon ~.
\label{nontarget}
\end{equation}

Step 1 of the algorithm, using an oracle query, flips ancilla-2 when the joint
register state is $\vert{t_j}\rangle$. In step 2, we measure ancilla-2. If
the outcome is $1$ then we stop the algorithm, because the register is in
the target state. The probability of getting $1$ is $\sin^{2}\alpha/2=f/2$;
we have effectively put an upper bound of $1/2$ on the success probability
using ancilla-1. If the outcome is $0$, which has probability $1/N^2$, then
the joint state is $\vert{t^{\prime}_j}\rangle$. In this case, we apply the
joint diffusion operation using the joint source state $\vert{s_j}\rangle
\equiv\vert{0}\rangle\vert{s}\rangle$. The joint diffusion operation is
a reflection about $(H\otimes U)|0\rangle|s\rangle\equiv U_j|s_j\rangle$,
in the two-dimensional Hilbert space spanned by $\vert{t_j}\rangle$ and
$\vert{t^{\prime}_j}\rangle$ as shown in Fig.3. The state $U_j\vert{s_j}\rangle$
makes an angle $\alpha_j$, defined by $\sin^2\alpha_j = \sin^2\alpha/2$,
with the state $\vert{t^{\prime}_j}\rangle$. So reflecting $\vert{t^{\prime}_j}
\rangle$ about $U_j\vert{s_j}\rangle$ gives us a state that makes an angle
$2\alpha_j$ with $\vert{t^{\prime}_j}\rangle$. Its component in
$\vert{t^{\prime}_j}\rangle$-direction is $\cos2\alpha_j = 1-2\sin^2\alpha_j
= 1-\sin^2\alpha = \epsilon$. Thus the joint diffusion operation produces
the final state
\begin{equation}
\vert{\psi_f}\rangle= U_{j} I_{s_{j}} U_{j}^{\dagger}|t^{\prime}_{j}\rangle=
\sqrt{1-\epsilon^{2}}\vert{t_{j}}\rangle+\epsilon\vert{t^{\prime}_{j}}
\rangle ~.
\end{equation}
After measuring $\vert{\psi_f}\rangle$, the probability of getting
$\vert{t^{\prime}_j}\rangle$ is $\epsilon^2$, so the total probability
of getting $\vert{t^{\prime}_j}\rangle$ after one iteration is
$N^{-2}\epsilon^{2}$. Iterating the algorithm will keep on decreasing this
probability by a factor of $\epsilon^2$ at each iteration, and after $q$
iterations it will become $N^{-2}\epsilon^{2q}$. So the net error probability
after $q$ iterations is (cf. Eq.(\ref{nontarget}))
\begin{equation}
\epsilon_{q} = N^{2}\epsilon(N^{-2}\epsilon^{2q}) = \epsilon^{2q+1} ~,
\end{equation}
which agrees exactly with the corresponding result for the Phase-$\pi/3$
search.

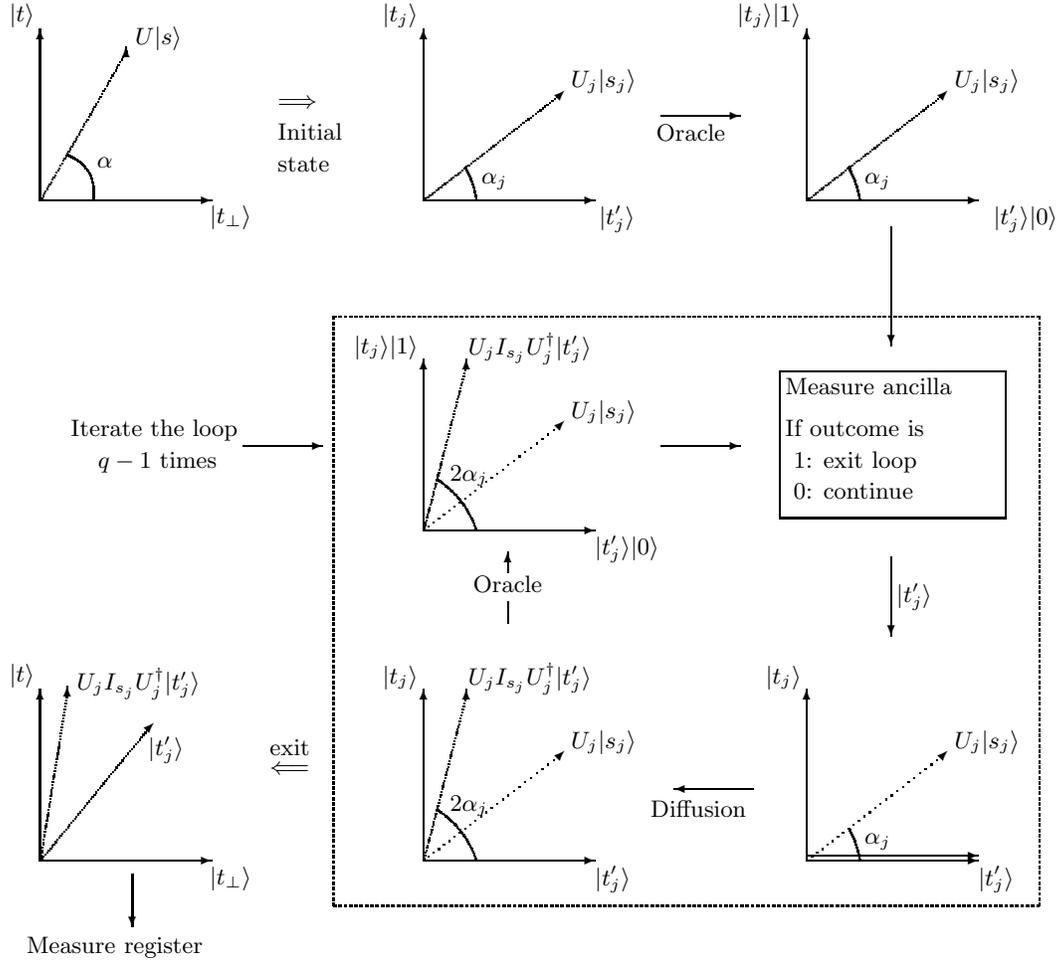
\begin{figure}[ptb]
\begin{picture}(500,390)(0,-40)
\renewcommand{\qbeziermax}{60}
\put(15,15){\vector(1,0){65}}
\put(15,15){\vector(0,1){65}}
\put(160,265){\vector(1,0){65}}
\put(15,265){\vector(1,0){65}}
\put(305,265){\vector(1,0){65}}
\put(305,15){\vector(1,0){65}}
\put(160,140){\vector(1,0){65}}
\put(160,15){\vector(1,0){65}}
\put(160,265){\vector(0,1){65}}
\put(15,265){\vector(0,1){65}}
\put(305,265){\vector(0,1){65}}
\put(305,15){\vector(0,1){65}}
\put(160,140){\vector(0,1){65}}
\put(160,15){\vector(0,1){65}}
\qbezier(15,265)(30,290.6)(47.5,321.3)
\put(47.5,321.3){\vector(0,1){2}}
\qbezier(160,265)(183.7,283.4)(211.4,304.8)
\put(211.4,304.8){\vector(1,1){2}}
\qbezier(305,265)(328.7,283.4)(356.4,304.8)
\put(356.4,304.8){\vector(1,1){2}}
\put(13,335){\makebox(0,0)[r]{$\vert{t}\rangle$}}
\put(95,263){\makebox(0,0)[tr]{$\vert{t_\perp}\rangle$}}
\put(51,331){\makebox(0,0)[tl]{$U\vert{s}\rangle$}}
\put(240,263){\makebox(0,0)[tr]{$\vert{t'_{j}}\rangle$}}
\put(158,335){\makebox(0,0)[r]{$\vert{t_{j}}\rangle$}}
\put(216,314.5){\makebox(0,0)[tl]{$U_{j}\vert{s_{j}}\rangle$}}
\put(400,263){\makebox(0,0)[tr]{$\vert{t'_{j}}\rangle\vert{0}\rangle$}}
\put(303,335){\makebox(0,0)[r]{$\vert{t_{j}}\rangle\vert{1}\rangle$}}
\put(361,314.5){\makebox(0,0)[tl]{$U_{j}\vert{s_{j}}\rangle$}}
\qbezier(25,282.3)(36.65,277.5)(35,265)
\put(36.65,277.5){$\alpha$}
\qbezier(175.8,277.25)(178.92,271.47)(180,265)
\put(182,271.47){$\alpha_{j}$}
\qbezier(320.8,27.25)(323.92,21.47)(325,15)
\put(327,21.47){$\alpha_{j}$}
\qbezier(320.8,277.25)(323.92,271.47)(325,265)
\put(327,271.47){$\alpha_{j}$}
\qbezier[25](305,15)(328.7,33.4)(356.4,54.8)
\put(356.4,54.8){\vector(1,1){2}}
\qbezier[25](160,15)(183.7,33.4)(211.4,54.8)
\put(211.4,54.8){\vector(1,1){2}}
\qbezier(160,15)(167.5,44.04)(176.25,77.93)
\put(176.25,77.93){\vector(0,1){2}}
\qbezier(160,140)(167.5,169.04)(176.25,202.93)
\put(176.25,202.93){\vector(0,1){2}}
\qbezier[25](160,140)(183.7,158.4)(211.4,179.8)
\put(211.4,179.8){\vector(1,1){2}}
\put(305,17){\vector(1,0){65}}
\put(295,145){\framebox(85,55)}
\put(297,198){\makebox(0,0)[tl]{Measure ancilla}}
\put(297,182){\makebox(0,0)[tl]{If outcome is}}
\put(297,170){\makebox(0,0)[tl]{ $1$: exit loop}}
\put(297,158){\makebox(0,0)[tl]{ $0$: continue}}
\put(370,13){\makebox(0,0)[tl]{$\vert{t'_{j}}\rangle$}}
\put(303,80){\makebox(0,0)[br]{$\vert{t_{j}}\rangle$}}
\put(361,64.5){\makebox(0,0)[tl]{$U_{j}\vert{s_{j}}\rangle$}}
\put(225,13){\makebox(0,0)[tl]{$\vert{t'_{j}}\rangle$}}
\put(158,80){\makebox(0,0)[br]{$\vert{t_{j}}\rangle$}}
\put(216,64.5){\makebox(0,0)[tl]{$U_{j}\vert{s_{j}}\rangle$}}
\put(225,138){\makebox(0,0)[tl]{$\vert{t'_{j}}\rangle\vert{0}\rangle$}}
\put(158,205){\makebox(0,0)[br]{$\vert{t_{j}}\rangle\vert{1}\rangle$}}
\put(216,189.5){\makebox(0,0)[tl]{$U_{j}\vert{s_{j}}\rangle$}}
\put(95,13){\makebox(0,0)[tr]{$\vert{t_\perp}\rangle$}}
\put(13,85){\makebox(0,0)[r]{$\vert{t}\rangle$}}
\put(126,-2){\dashbox(267,223)}
\put(177,203){\makebox(0,0)[bl]{$U_{j}I_{s_{j}}U_{j}^{\dagger}\vert{t'_{j}}\rangle$}}
\put(177,78){\makebox(0,0)[bl]{$U_{j}I_{s_{j}}U_{j}^{\dagger}\vert{t'_{j}}\rangle$}}
\put(105,300){$\Longrightarrow$}
\put(105,288){Initial}
\put(105,276){state}
\put(245,294){\makebox(0,0)[tl]{ Oracle}}
\put(250,297){\vector(1,0){30}}
\put(337,255){\vector(0,-1){45}}
\put(337,130){\vector(0,-1){30}}
\put(285,42){\vector(-1,0){30}}
\put(192,105){\line(0,1){8}}
\put(192,124){\vector(0,1){8}}
\put(250,172){\vector(1,0){30}}
\put(192,120){\makebox(0,0){Oracle}}
\put(246,38){\makebox(0,0)[tl]{Diffusion}}
\qbezier(165,34.36)(175.81,27.25)(180,15)
\put(170,34){$2\alpha_{j}$}
\put(170,159){$2\alpha_{j}$}
\qbezier(165,159.36)(175.81,152.25)(180,140)
\put(339,115){\makebox(0,0)[l]{$\vert{t'_{j}}\rangle$}}
\qbezier(15,15)(33.97,38.238)(56.11,65.35)
\qbezier(15,15)(19.74,44.62)(25.28,79.18)
\put(56.11,65.35){\vector(1,1){2}}
\put(25.28,79.18){\vector(0,1){2}}
\put(56,62){\makebox(0,0)[tl]{$\vert{t'_{j}}\rangle$}}
\put(29,76){\makebox(0,0)[bl]{$U_{j}I_{s_{j}}U_{j}^{\dagger}\vert{t'_{j}}\rangle$}}
\put(102,48){$\Longleftarrow$}
\put(102,55){exit}
\put(50,10){\vector(0,-1){20}}
\put(10,-20){Measure register}
\put(92,172){\vector(1,0){30}}
\put(90,174){\makebox(0,0)[br]{Iterate the loop}}
\put(82,162){\makebox(0,0)[br]{$q-1$ times}}
\end{picture}
\caption{Step-by-step quantum state evolution in our $q$-iteration algorithm.
The double arrows relate the equivalent states in the original search space
(left) and the joint search space (right).}
\end{figure}

The above analysis allows us to also deduce the following features:
\newline(1) $\epsilon=1$ can be made a fixed point of the algorithm, instead
of $\epsilon=0$, by effectively interchanging the roles of $|t\rangle$ and
$|t_{\perp}\rangle$. This is achieved by flipping ancilla-2, only when the
joint state is $|1\rangle|t_{\perp}\rangle$. Then the probability of finding
the register in the state $|t\rangle$, after $q$ iterations, becomes
$(1-\epsilon)^{2q+1}$. Note that the same behavior can be obtained in
the Phase-$\pi/3$ search, by replacing either $R_{t}^{\pi/3}$ with
$R_{t}^{-\pi/3}$ or $R_{s}^{\pi/3}$ with $R_{s}^{-\pi/3}$. This $\epsilon=1$
fixed point can be useful in situations where certain target states are to be
avoided, e.g. in collision problems.
\newline(2) When we use fixed point quantum search to locate the target state
in a database, the initial error probability is $\epsilon=1-f$. The number
of oracle queries required to reduce this probability to $o(1)$ obeys
\begin{equation}
(1-f)^{2q+1} = o(1) ~~\mathop{\longrightarrow}\limits^{f~small}~~ e^{-(2q+1)f}
= o(1) ~.
\end{equation}
Thus we need $q=O(1/f)$ oracle queries to find the target state reliably.
This scaling of fixed point quantum search is clearly inferior to the
$O(1/\sqrt{f})$ scaling of the quantum search algorithm~\cite{grover96}.
Still, as discussed in the introduction, fixed point quantum search
can be useful in situations where $f$ is unknown and large.
\newline(3) A practical criterion for stopping the iterative algorithm would
be that the error probability becomes smaller than some predetermined
threshold $\epsilon_{\mathrm{th}}$. Provided we have an upper bound
$\epsilon\le\epsilon_{\mathrm{up}}<1$, we can guarantee convergence by
choosing $\epsilon_{\mathrm{up}}^{2q+1}\le\epsilon_{\mathrm{th}}$.
In the above algorithm $2q+1$ can take all odd positive integer values,
which is less restrictive than the values of the form $3^n$ allowed by
the Phase-$\pi/3$ search.
\newline(4) Depending on the outcome of the ancilla-2 measurements, the
algorithm can exit the iteration loop before completing it. So the average
number of queries is always less than the maximum number of iterations $q$.
This is better than the Phase-$\pi/3$ search, where all the queries required
by the number of iterations must be executed.
\newline(5) It is possible to stick to unitary operations throughout the
algorithm and postpone measurement till the very end. In such a scenario,
the unmeasured ancilla-2 has to control the diffusion operation and all the
subsequent iterations (i.e. they are executed only when ancilla-2 is in the
$|0\rangle$ state), and it cannot be reused in the iteration loop. We need
a separate ancilla-2 for every oracle query, to ensure that once the states
$|t_{j}\rangle$ and $|t^{\prime}_{j}\rangle$ are separated by an oracle query
in an iteration, they are not superposed again by subsequent iterations.
The whole set of $q$ ancilla-2 can be measured after the iteration loop,
in a sequence corresponding to the iteration number, to determine the target
state. In this version, the unitary transformation is different for each
iteration (because each iteration involves a different ancilla-2 and
different controls), and the unitary iterations converge to a fixed point.

\section{Quantum Searching amidst Uncertainty}

The original quantum search algorithm is known to be the best possible
algorithm for exhaustive searching~\cite{bbbv,zalka}, therefore no
algorithm will be able to improve its performance. However, for applications
other than exhaustive searching for a single item, we demonstrate that
suitably modified algorithms may indeed provide better performance.

Consider the situation where a large fraction of the states are marked, but
the precise fraction of marked states is not known. The goal is to find a
single marked state with as high a probability as possible in a single query.
For concreteness, say the marked fraction $f$ is uniformly distributed
between 75\% and 100\%. In that case, as shown below, the probability of
error for the fixed point search algorithm is approximately one fifth of
that for the best (possible) classical algorithm, and approximately one
fifteenth of that for the best (known) quantum algorithm.

\textbf{Classical:}
The best classical algorithm is to select a random state and see if it
is the target state (one query). If yes, return this state; if not, pick
another random state and return that. The probability of error is equal
to that of not getting a single target state in two random picks,
i.e. $\epsilon^2=(1-f)^2$, which lies in the interval $[0,1/16]$.
The overall error probability is approximately 2.1\%.

\textbf{Quantum:}
The best quantum search based algorithm available in the literature, for
this problem, is by Younes {\it et al.}~\cite{younes}. Using $q$ queries,
it finds the target state with a probability $f \left( \frac{\sin^2(q+1)\beta}
{\sin^{2}\beta}+\frac{\sin^{2}q\beta}{\sin^{2}\beta}\right)$, where
$\cos\beta=\epsilon$ (see Eq.(59) of Ref.~\cite{younes}). When $q=1$,
the success probability becomes $f\left(1+4(1-f)^{2}\right)$, which lies
in the interval $[15/16,1]$. The overall error probability is approximately
5.7\%, which is worse than that for the best classical algorithm.

\textbf{Fixed point search:}
If we use the transformation $O(\pi/3,\pi/3)$, with $U$ being the
Walsh-Hadamard transform and the source state $|s\rangle$ being the
$|\overline{0}\rangle$ state, then the probability of being in a non-target
state after one query becomes $\epsilon^3=(1-f)^3$, which lies in the
interval $(0,1/64)$. The overall error probability is approximately 0.4\%,
far superior to the two cases described above.

\begin{figure}[ptb]
\begin{center}
\epsfxsize=15truecm
\centerline{\epsfbox{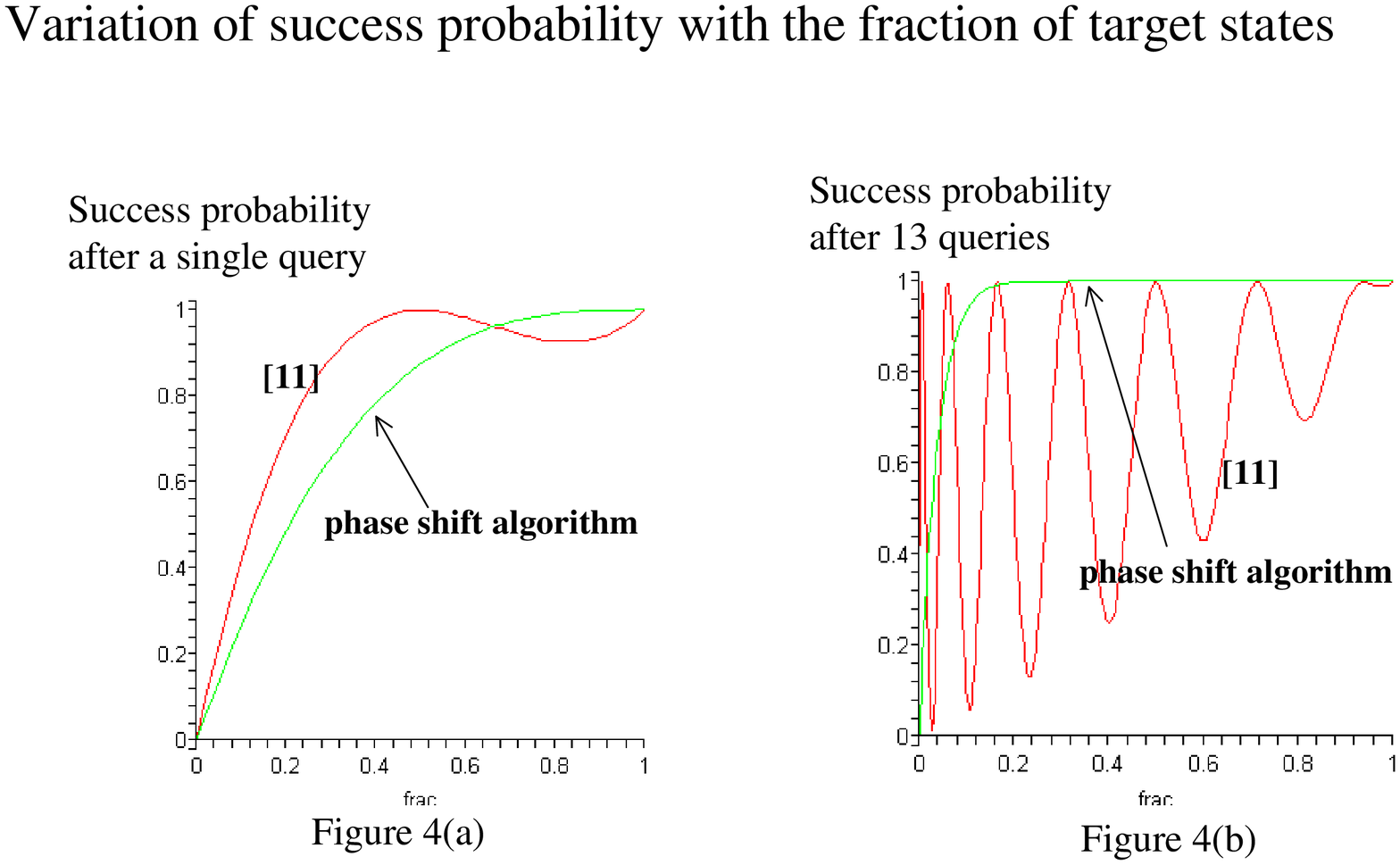}}
\vspace{-1truecm}
\caption{Comparison of the performance of the Phase-$\pi/3$ search algorithm
with that of Ref.\cite{younes}, when the fraction of marked states, $f$,
varies between $0$ and $1$: (a) after one query, (b) after 13 queries.}
\end{center}
\end{figure}

How the quantum algorithms perform is graphically illustrated in Fig.4.
For the particular distribution of $f$ considered above, Fig.4(a) shows
that the graph of the fixed point search algorithm lies entirely above
the graph of Ref.~\cite{younes}. The difference between the two becomes
even more dramatic if we consider multiple query algorithms, as in Fig.4(b).
As mentioned earlier, for the type of problems discussed in this section,
the fixed point algorithm is the best possible quantum algorithm~\cite{jai}.

\section{Conclusion}

The variant of quantum search discussed in this article supplements the
original search algorithm, by providing a scheme that permits a fixed point
and hence moves towards the target state in a directed way. The fixed point
property makes the quantum search robust, and naturally leads to schemes
for correction of systematic errors~\cite{reichardt}.

We have presented two implementations of the fixed point quantum search,
both optimal in the worst-case scenario. At first sight, requirements of
monotonic convergence and unitary quantum evolution appear to be in conflict.
But the conflict has been avoided by a recursive approach in the Phase-$\pi/3$
search algorithm and by irreversible projections in the measurement based
search algorithm. Between the two, the measurement based search algorithm
is better behaved than the Phase-$\pi/3$ search, because (a) it allows all
positive integer number of oracle queries instead of a restricted set of
integers, and (b) its average-case requirement of number of oracle queries
is smaller.

Deeper insights in to the mechanisms of these algorithms are still desirable.
For instance, what is so special about phase shift of $\pi/3$, and how does
the change of phase shift (from $\pi$ to $\pi/3$) convert the amplitude
amplification algorithm to something totally different? Can one improve
the algorithms in the non-asymptotic region by making the phase shifts
(or rotation angles) a function of the query sequence, in a manner
reminiscent of local adiabatic quantum algorithms? Such questions are
definitely worth exploring.

The potential applications of these algorithms are in problems with small
but unknown initial error probability and limited number of oracle queries,
where they assure monotonic power-law convergence. For such problems, the
fixed point algorithms are better than the best classical search algorithms
by a factor of 2. Applications to pattern recognition and associative memory
recall are under investigation.

\end{document}